\documentclass[3p]{elsarticle}

\usepackage{algorithm}
\usepackage{algorithmic}
\usepackage{graphicx} % figure
\usepackage{booktabs}
\usepackage{amssymb}
\usepackage{amsmath}
\usepackage{bm}
\usepackage{multirow}
\usepackage{colortbl}
\usepackage{color}
\usepackage{array}
\begin{document}

\begin{frontmatter}
	
	\title{A Generative Node-attribute Network Model for Detecting Generalized Structure}% Force line breaks with \\
	
	\author[lab1]{Wei Liu}
	
	\author[lab1]{Zhenhai Chang \corref{mycorrespondingauthor}}
	\cortext[mycorrespondingauthor]{Corresponding author}
	\ead{changzhenhai2012@163.com}
	
	\author[lab2,lab3]{Caiyan Jia \corref{mycorrespondingauthor}}%
	\cortext[mycorrespondingauthor]{Corresponding author}
	\ead{cyjia@bjtu.edu.cn}

	\author[lab2,lab3]{Yimei Zheng}
	
	\address[lab1]{School of Mathematics and Statistics, Tianshui Normal University, Gansu, China}
	\address[lab2]{School of Computer and Information Technology, Beijing Jiaotong University, Beijing, China}
	\address[lab3]{Beijing Key Lab of Traffic Data Analysis and Mining, Beijing, China}

	\date{\today}% It is always \today, today,
	
	\begin{abstract}
		Exploring meaningful structural regularities embedded in networks is a key to understanding and analyzing the structure and function of a network. The node-attribute information can help improve such understanding and analysis. However, most of the existing methods focus on detecting traditional communities, i.e., groupings of nodes with dense internal connections and sparse external ones. In this paper, based on the connectivity behavior of nodes and homogeneity of attributes, we propose a principle model (named GNAN), which can generate both topology information and attribute information. The new model can detect not only community structure, but also a range of other types of structure in networks, such as bipartite structure, core-periphery structure, and their mixture structure, which are collectively referred to as generalized structure. The proposed model that combines topological information and node-attribute information can detect communities more accurately than the model that only uses topology information. The dependency between attributes and communities can be automatically learned by our model and thus we can ignore the attributes that do not contain useful information. The model parameters are inferred by using the expectation-maximization algorithm. And a case study is provided to show the ability of our model in the semantic interpretability of communities. Experiments on both synthetic and real-world networks show that the new model is competitive with other state-of-the-art models.
	\end{abstract}
	

\end{frontmatter}

	\section{\label{Introduction}Introduction}
	Many systems in the real world can be simplified as networks, where each node (vertex) represents an individual and an edge exists between two nodes if the two corresponding individuals interact in some way. Examples include friendships in social networks of interactions among people \citep{HE202134,West18355}, molecular bindings in biological networks of molecules \citep{SantoliniE6375,2020Robustness,Gro2019Representing}, and web hyperlinks in the World Wide Web \citep{barabasi2000scale}. In the past, most studies have only treated networks as unadorned sets of nodes and their links \citep{girvan2002community,5fortunato2009community}. In recent years, most network data, however, are accompanied by the contents that describe the properties of nodes. For example, a user on Twitter and users whom he/she follow represents relationships while Twitter lists and tweets they post describe the profile of the user. Such networks represented by the semantic contents combining with links (also called attribute information and topology information respectively) are referred as attributed networks or attributed graphs \citep{li2018community,2020Network,ren2021block}. 
	
	As a network structure sheds light on the behavior of a system in a way, a large number of studies have been devoted to the detection of community structures in networks \citep{5fortunato2009community,chunaev2020community,2009Communities}. However, most of these studies have focused on methods of discovering traditional communities, i.e., groupings of nodes with dense internal connections and sparse external ones. In fact, with the emergence of various complex networks in different fields, including social, information, biological and physical sciences, different types of network structures are discovered and studied. Examples include bipartite structure, core-periphery structure, and their mixture structure, etc. Here traditional communities and other types of communities are referred to as generalized structures\citep{matias2014modeling,2015Generalized}.
	
	In recent years, some methods have been proposed for the exploration of structures contained in node-attribute networks \citep{chunaev2020community,2015Clustering}, which roughly fall into two categories according to how to use attributes: the methods that use full attribute space \citep{bojchevski2018bayesian,chang2019generative,jin2019detecting,chen2016network,chai2013combining} and the methods that explore subspaces of attributes \citep{gunnemann2011db,gunnemann2013spectral,perozzi2014focused,wu2018mining,gunnemann2013efficient,huang2017joint,chen2020attributed}. The former fuses structure and all available attributes to improves community detection quality, the latter believes that part attributes are related to obtaining good-quality communities. One of the former subclasses is the probabilistic model-based methods, which generate links and node attributes through a joint probability function and the model parameters. The generative models can be further classified into two categories in terms of types of network structure: one mainly detects traditional communities, and the other detects generalized communities, including traditional communities. For example, Yang et al. \citep{yang2009combining} combined a popularity-based conditional link Model PCL with a discriminative content (DC) model for community detection (termed PCL\_DC). By introducing node productivity, Yang et al. \citep{yang2010directed} further developed a popularity and productivity link model PPL, and the corresponding united model was called PPL\_DC. However, both PCL\_DC and PPL\_DC only detect traditional network structure since they assume that nodes in the same community have more opportunities to link each other. On the contrary, due to Newman’s mixture model (NMM) \citep{newman2007mixture}, the method BNPA proposed by Chen et al. \citep{chen2016network} can detect generalized network structure. Thanks to the block structure assumption of SBM \citep{holland1983stochastic}, the models of Chai et al. \citep{chai2013combining}, He et al. \citep{he2017joint}, and Chang et al. \citep{chang2019generative} can discover generalized network structure. However, Chai et al. \citep{chen2016network} also used a DC model for attributes so that the proposed model PPSB\_DC was not good at semantic interpretability because the model only learned a weight vector of attributes for communities. He et al. \citep{he2017joint} developed a model NEMBP that had good semantic interpretability because the relationship between a community and its corresponding attributes was characterized. However, NEMBP needed to specify the number of both topics and communities in advance. PSB\_PG \citep{chang2019generative} also had good semantic interpretability but had nonlinear complexity in a naive EM algorithm, which means the method works well for networks of moderate size. 
	
	Here we propose a principle generative model to address the above problems. Firstly, based on the connectivity behavior of nodes, a model that can generate topology information is proposed. Secondly, based on the homogeneity of attributes, a model that can generate attribute information is developed. Finally, the two models are united together by sharing the latent locations of nodes and a generative node-attribute network model (named GNAN) is formed naturally. The main contributions of the proposed method are summarized as follows.
	\begin {itemize}
	\item The proposed model that combines topological information with attribute information can detect communities more accurately than the model that only uses topology information, which means the node attributes are effectively utilized and can complement the network structure.
	
	\item   The new model GNAN can classify the nodes of a network into groups such that the link patterns of each group are similar in some sense. Therefore, the model can detect not only traditional communities, but also a range of other types of structures in networks, such as bipartite structure, core-periphery structure, and their mixture structure.
	
	\item The dependency between attributes and communities can be automatically learned by our model and thus we can ignore the attributes that do not contain useful information. A case study is provided to show the ability of our model in the semantic interpretability of communities. And experiments on both synthetic and real-world networks show that the new model is competitive with other state-of-the-art models.
	\end {itemize}
	
	The rest of the paper is organized as follows. In Sect.II, a generative node-attribute network model is described. In Sect.III, the model parameters are inferred and the corresponding algorithm is designed. In Sect.IV, the new algorithm is evaluated and compared with some related methods on both synthetic and real-world networks. Finally, In Sect.V, we derive the conclusions.

	\section{\label{GNAN}GNAN: A generative node-attribute network model}
	Let $ G(V, E, X) $ be a mathematical formalization for a network, where $ V = \left\{ {1,2, \cdots ,N} \right\} $ is the set of nodes (vertices), $ E = \left\{ {e_1 ,e_2 , \cdots ,e_M } \right\} $ is the set of links (edges), $ N $ and $ M $ are the number of nodes and links, respectively. $ X = \left( {x_{ik} } \right)_{N \times K} $ is a node-attribute matrix of a network and represents the attribute information contained in a network data set, $ K $ is the dimension of the node attributes, $ x_{ik}  = 1 $ means node $ i $ has the $ k $th attribute, or 0 otherwise. The topology information of a network is represented by an adjacent matrix $ A = \left( {a_{ij} } \right)_{N \times N} $, where $ a_{ij}  = 1 $ if a link between the pair of nodes $ (i, j) \in E$, or 0 otherwise. Suppose $ V_1, V_2, \cdots , V_C $ are the communities embedded in a network, and $	\bigcup\limits_{r = 1}^C {V_r }  = V $.
	
\begin{figure}[htbp]
		\centering
		\includegraphics[width=14.5cm]{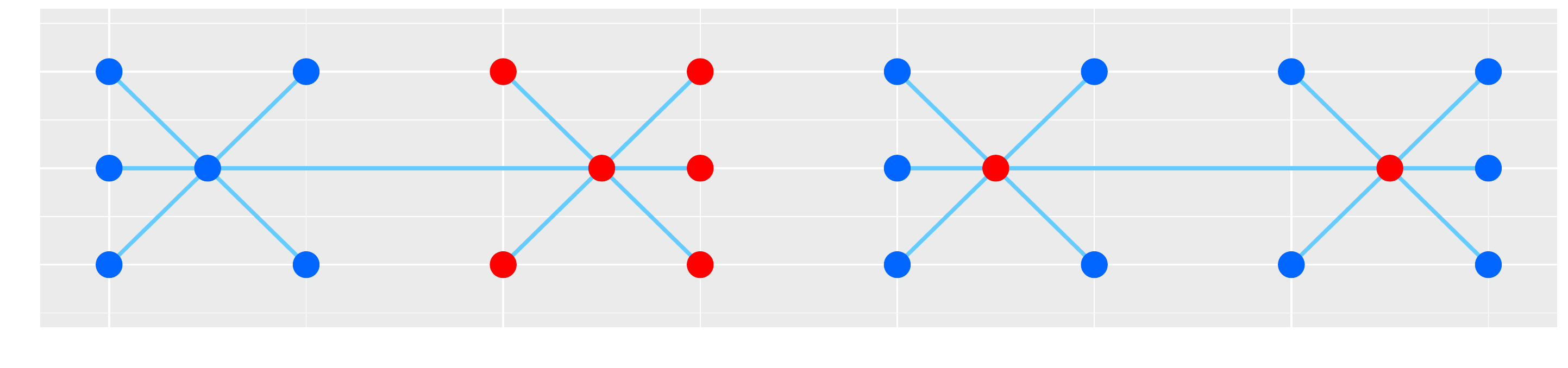}
		\caption{A toy network with 2 communities represented by colors (blue or red). The left panel means that a community is a set of nodes that share a large number of connections, i.e., a traditional community. The right panel means that a community is a set of nodes that share the same connectivity behavior, i.e., a generalized community. Of course, the situation of the left panel can also be regarded as an example based on the connectivity behavior of nodes.}\label{fig0}
	\end{figure}

	\textbf{Modeling the links of a network} In order to enable the topology structure generated by the new model to form a wide range of network structures, we consider introducing a group of parameters that characterize the connectivity behavior of nodes. A community is a set of nodes that share the same connectivity behavior, which is in line with our intuition (See a toy example in FIG.\ref{fig0} \citep{matias2014modeling}).
	
	In this paper, we adopt a parameter matrix $ \Theta  = \left( {\theta _{rj} } \right)_{C \times N} $ to measure the connectivity behavior of nodes, where the entry $ \theta_{rj} $ represents the probability that any particular node in community $ V_r $ sends an edge to node $ j $, and $ \sum\limits_{j = 1}^N {\theta _{rj} }  = 1 $, which is also used in NMM \citep{newman2007mixture}. Thus, $ \theta_{rj} $ characterizes the preference of nodes in community $ V_r $ about which other nodes they like to link to. These preferences form the communities in which nodes have similar patterns of connection to others. To generate an expected link between a pair of nodes $ (i, j) $, another parameter matrix $ {\rm T} = \left( {\tau _{ir} } \right)_{N \times C} $ is introduced to our model, where the entry $ \tau _{ir} $ represents the probability that a node $ i $ falls into community $ V_r $, and $ \sum\limits_{r = 1}^C {\tau _{ir} }  = 1 $. Based on both parameters $ \tau _{ir} $ and $ \theta_{rj} $, an expected link between node pairs $ (i, j) $ through $ V_r $ is $ \hat a_{ij,r}  = \tau _{ir} \theta _{rj} $. Summing over communities $ V_r $, the expected number of links between a pair of nodes $ (i, j) $ is $ \hat a_{ij}  = \sum\limits_{r = 1}^C {\tau _{ir} \theta _{rj} } $. Suppose directed links are placed independently between node pairs with probabilities that are Poisson distribution, the likelihood of topology information is
	\[
	\Pr \left( {A|{\rm T},\Theta } \right) = \prod\limits_{i,j = 1}^N {\frac{{\hat a_{ij}^{a_{ij} } }}{{a_{ij} !}}\exp \left\{ { - \hat a_{ij} } \right\}}  = \prod\limits_{i,j = 1}^N {\frac{{\left( {\sum\limits_{r = 1}^C {\tau _{ir} \theta _{rj} } } \right)^{a_{ij} } }}{{a_{ij} !}}\exp \left\{ { - \sum\limits_{r = 1}^C {\tau _{ir} \theta _{rj} } } \right\}}.  \eqno(1) 
	\]
	
	\textbf{Modeling the attributes of nodes in a network} In order to automatically learn the dependency between attributes and communities, a parameter matrix $ \Phi  = \left( {\phi _{rk} } \right)_{C \times K} $ is introduced to the model of generating node attributes, where the entry $ \phi _{rk} $ represents the probability that a community $ V_r $ has the $ k $th attribute, and $ \sum\limits_{k = 1}^K {\phi _{rk} }  = 1 $. Therefore, a node $i$ in community $V_r$ possessing $k$th attribute can be represented as $ \hat x_{ik,r}  = \tau_{ir} \phi _{rk} $. Summing over all communities $V_r$, the expected propensity of a node $i$ possessing $k$th attribute is $ \hat x_{ik}  = \sum\limits_r {\tau_{ir} \phi _{rk} } $. Suppose $x_{ik} (k=1,2,\dots,K)$ is independent and identically distributed, we have 
	\[
	\Pr \left( {X|{\rm T},\Phi } \right) = \prod\limits_{i = 1}^N {\prod\limits_{k = 1}^K {\frac{{\hat x_{ik}^{x_{ik} } }}{{x_{ik} !}}\exp \left\{ { - \hat x_{ik} } \right\}} }  = \prod\limits_{i = 1}^N {\prod\limits_{k = 1}^K {\frac{{\left( {\sum\limits_{r = 1}^C {\tau _{ir} \phi _{rk} } } \right)^{x_{ik} } }}{{x_{ik} !}}\exp \left\{ { - \sum\limits_{r = 1}^C {\tau _{ir} \phi _{rk} } } \right\}} }.  \eqno(2) 
	\]
	
	In effect, $ \phi _{rk} $ characterizes the preference of which attributes nodes in community $ V_r $ possess, which is similar to the parameter $ \theta_{rj} $ in some sense. The preferences define communities in which nodes have the same attributes (named attribute communities here to discriminate the communities formed by $ \theta_{rj} $). We can order the attributes learned by $ \theta_{rj} $ and easily choose some important attributes for a certain community. These selected attributes may come from the same topic or different ones if the attributes have topics, but they are shared simultaneously by nodes in community $ V_r $, we call such property homogeneity of attributes. The shared attributes naturally build up the semantic interpretation for each community. Note that the constraints are added to the columns of the parameter $ \Phi $, which means that an attribute can maintain a close relationship with multiple communities at the same time. Therefore, there may not be a one-to-one relationship between communities and attribute topics if the attributes have topics.
	
	\textbf{GNAN} By sharing the latent position of nodes, the joint likelihood function for generating node-attribute network can be described as follows:
	\[
	\begin{array}{l}
	\Pr \left( {A{\rm{,}}X|{\rm T},\Theta ,\Phi } \right) = \Pr \left( {A|{\rm T},\Theta } \right)\Pr \left( {X|{\rm T},\Phi } \right) \\ 
	\;\;\;\;\;\;\;\;\;\;\;\;\;\;\;\;\;\;\;\;\;\;\;\;\;\;\;\; = \prod\limits_{i,j = 1}^N {\frac{{\left( {\sum\limits_{r = 1}^C {\tau _{ir} \theta _{rj} } } \right)^{a_{ij} } }}{{a_{ij} !}}\exp \left\{ { - \sum\limits_{r = 1}^C {\tau _{ir} \theta _{rj} } } \right\}}  \\ 
	\;\;\;\;\;\;\;\;\;\;\;\;\;\;\;\;\;\;\;\;\;\;\;\;\;\;\;\;\;\;\;\;\;\;\; \times \prod\limits_{i = 1}^N {\prod\limits_{k = 1}^K {\frac{{\left( {\sum\limits_{r = 1}^C {\tau _{ir} \phi _{rk} } } \right)^{x_{ik} } }}{{x_{ik} !}}\exp \left\{ { - \sum\limits_{r = 1}^C {\tau _{ir} \phi _{rk} } } \right\}} }, \\   
	\end{array}  \eqno(3) 
	\]
	where `` $ \times $ " represents multiplication. Under the assumption of the sparsity of links and attributes, a unified Poisson distribution likelihood function makes the model both reasonable and easy to calculate.

	\section{A proposed algorithm for detecting generalized structure with GNAN}
	In this section, firstly, the parameters $ {\rm T} $ and $ \Phi $ in model Eq.(3) are inferred by the EM algorithm. The learned parameter $ {\rm T} $ can help us to derive the network structures embedded in a network. The inferred parameter $ \Phi $ represents the dependency of communities and attributes, helping to explain why these nodes come together. Then, based on the inferred parameters, an algorithm for GNAN is designed. 
	
	Because the parameters in model Eq.(3) are related to the potential position of nodes that cannot be observed (i.e., a latent variable), it is difficult to directly estimate them. The EM algorithm can conveniently handle this type of parameter estimation problem with latent variables. Considering the logarithm of the model Eq.(3), ignoring constants and terms independent of parameters and latent variables, we have
	\[
	L\left( {{\rm T},\Theta ,\Phi } \right) = \sum\limits_{i.j = 1}^N {\left[ {a_{ij} \log \left( {\sum\limits_{r = 1}^C {\tau _{ir} \theta _{rj} } } \right) - \sum\limits_{r = 1}^C {\tau _{ir} \theta _{rj} } } \right]}  + \sum\limits_{i = 1}^N {\sum\limits_{k = 1}^K {\left[ {x_{ik} \log \left( {\sum\limits_{r = 1}^C {\tau _{ir} \phi _{rk} } } \right) - \sum\limits_{r = 1}^C {\tau _{ir} \phi _{rk} } } \right]} }.  \eqno(4) 
	\]
	
	From Jensen's inequality, the lower bound of the log-likelihood Eq.(4) is as follows:
	\[
	\bar L\left( {{\rm T},\Theta ,\Phi } \right) = \sum\limits_{ijr} {a_{ij} q_{ij,r} \log \frac{{\tau _{ir} \theta _{rj} }}{{q_{ij,r} }}}  - \sum\limits_{ijr} {\tau _{ir} \theta _{rj} }  + \sum\limits_{ikr} {x_{ik} h_{ik,r} \log \frac{{\tau _{ir} \phi _{rk} }}{{h_{ik,r} }}}  - \sum\limits_{ikr} {\tau _{ir} \phi _{rk} },    \eqno(5) 
	\]
	where
	\[
	q_{ij,r}  = \frac{{\tau _{ir} \theta _{rj} }}{{\sum\limits_{r = 1}^C {\tau _{ir} \theta _{rj} } }},\;h_{ik,r}  = \frac{{\tau _{ir} \phi _{rk} }}{{\sum\limits_{r = 1}^C {\tau _{ir} \phi _{rk} } }}.   \eqno(6) 
	\]
	are the expected probabilities of a node pair $ (i, j) $ in community $ V_r $ to be linked and those of nodes $i ( \in V_r ) $ possessing $k$th attribute, respectively. By using the Lagrange multiplicator method, we can obtain the estimates of $ {\rm T} $, $\Theta$, and $\Phi$ that maximize the lower bound $ \bar L\left( {{\rm T},\Theta ,\Phi } \right) $ in Eq.(5) in the following. 
	\[
	\tau _{ir}  = \frac{{\sum\limits_j {a_{ij} q_{ij,r} }  + \sum\limits_k {x_{ik} h_{ik,r} } }}{{\sum\limits_{jr} {a_{ij} q_{ij,r} }  + \sum\limits_{kr} {x_{ik} h_{ik,r} } }},\;\theta _{rj}  = \frac{{\sum\limits_i {a_{ij} q_{ij,r} } }}{{\sum\limits_{ij} {a_{ij} q_{ij,r} } }},\;\phi _{rk}  = \frac{{\sum\limits_i {x_{ik} h_{ik,r} } }}{{\sum\limits_{ik} {x_{ik} h_{ik,r} } }}.  \eqno(7) 
	\]
	
	See the appendix for a detailed derivation. Eqs.(6,7) build up our EM algorithm for GNAN, which will converge \citep{wu1983convergence}. 

	\begin{algorithm}[H]
	\caption{EM algorithm for GNAN}
	\renewcommand{\algorithmicrequire}{ \textbf{Input:}} %Use Input in the format of Algorithm
	\renewcommand{\algorithmicensure}{ \textbf{Output:}} %UseOutput in the format of Algorithm
	\label{Al:01}
	\begin{algorithmic}[1]
		\REQUIRE~~\\
		the adjacency matrix $A$, the attribute matrix $ X $\\
		the number of communities $C$, \\
		the maximum iteration $I_{T}$, and the threshold $\epsilon$ \\
		\ENSURE ~~\\
		the inferred parameters $ {\rm T},\Theta, \Phi $
		\vspace{2mm}
		\STATE Initialize $ {\rm T^{(0)}},\Theta^{(0)}, \Phi^{(0)} $ by Eq.(7).
		\STATE Compute lower bound $ \bar L\left( {{\rm T^{(0)}},\Theta^{(0)} ,\Phi^{(0)} } \right) $ by Eq.(5).
		\FOR{$t=1:I_{T}$} 
		\vspace{1mm}
		\STATE E-step: Compute $ q_{ij,r},\; h_{ik,r} $ by Eq.(6).
		\vspace{1mm}
		\STATE M-step: Compute $ {\rm T^{(t)}},\Theta^{(t)}, \Phi^{(t)}  $ by Eq.(7).
		\vspace{1mm}
		\STATE Compute lower bound $ \bar L\left( {\rm T^{(t)}},\Theta^{(t)}, \Phi^{(t)} \right) $ by Eq.(5).
		\vspace{1mm}	
		\IF{ $ \left| \bar L\left( {\rm T^{(t)}},\Theta^{(t)}, \Phi^{(t)} \right)- \bar L\left( {\rm T^{(t-1)}},\Theta^{(t-1)}, \Phi^{(t-1)}  \right)  \right| < \epsilon $ or $ t=I_{T} $}
		\vspace{1mm}
		\STATE $ \hat{\rm T} = {\rm T^{(t)}} ,\; \hat{\Theta}  = \Theta ^{(t)} ,\; \hat{\Phi}  = \Phi ^{(t)} $; STOP;
		\ENDIF
		\ENDFOR
	\end{algorithmic}
\end{algorithm}
	
	Algorithm 1 (called GNAN for simplicity) will converge to a local optimum of the likelihood. Therefore, there usually have different solutions from different starting points. For the robustness of our results, the initialization about parameters $ {\rm T},\Theta, \Phi $ is generated from a uniform distribution over [0.5-$ \zeta $,0.5+$ \zeta $], where $ \zeta $ is a random perturbation. In the following experiments, the maximum iteration $I_{T}=500$, and the threshold $\epsilon=10^{-6}$.

	The time complexity of the new algorithm is dominated by updating $ q_{ij, r}, h_{ik, r} $ in step 4 and $ \tau _{ir}, \theta _{rj}, \phi _{rk} $ in step 5. Updating them for all nodes takes $ O(MC) $ and $ O(NKC) $ operations, which is linear in the number of communities $C$, the number of links $M$,  the dimension of each node attribute $ K $, and the number of nodes $N$. Therefore, the total time complexity of the algorithm is $O((MC+NKC)I_T)$, where $I_T$ is the number of iterations. The memory of the method outlined above is dominated by updating $ q_{ij, r} $ and $ h_{ik, r} $ in step 4. The space required to store $ q_{ij, r} $ is $ O(MC) $ while the $ h_{ik, r} $ is $ O(NKC) $. Therefore, the whole memory use of the algorithm is $ O(MC+NKC) $, which is linear in $ N, M, C $, and $ K $. Obviously, the complexity of the GNAN is lower than that of the PSB\_PG ($ O(MC^2+NKC^2) $).

	\section{Experiments}
     Firstly, the ability of the new method to discover community and node-attribute information was shown on artificial networks. Then, we applied our method to a real friendship network to show the semantic interpretation of communities in a case study. Finally, our new model GNAN was evaluated on synthetic and real-world networks with a range of known network structures in comparison with 4 state-of-the-art methods. The metric NMI (Normalized Mutual Information) \citep{danon2005comparing} was used here to evaluate an algorithm running on a network with ground-truth communities.
	
	\textbf{NMI.} Suppose $V = \left( {V_{1} ,V_{2} , \cdots ,V_{C} } \right)$ are true communities in a network, $V^{*} = \left( {V_{1}^{*} ,V_{2}^{*} , \cdots ,V_{C}^{*} } \right)$ are inferred communities. NMI is defined as follows
	\[
	NMI(V ,V^{*} ) = \frac{\displaystyle {-2 \sum\limits_{i = 1}^C {\sum\limits_{j = 1}^C {N_{ij} \log \frac{{NN_{ij} }}{{N_i^{V } N_j^{V^{*} } }}} } }}{{\sqrt { \displaystyle \left( {\sum\limits_{i = 1}^C {N_i^{V} \log \frac{{N_i^{V} }}{N}} } \right)\left( {\sum\limits_{j = 1}^C {N_j^{V^{*} } \log \frac{{N_j^{V^{*} } }}{N}} } \right)} }}, \eqno(8)
	\]
	where $C$ is the number of communities in a network; $N$ is the number of nodes; $N_{ij}$ is the number of nodes in the true community $V_i$ that are assigned to the inferred community $V_{j}^{*}$; ${N_i^{V} }$ is the number of nodes in the true community $V_i$; ${N_j^{V^{*} } }$ is the number of nodes in the inferred community $V_{j}^{*}$. A larger $NMI$ means a better partition.

	\subsection{Performance on synthetic networks}
	In this section, the performance of the GNAN algorithm was tested on artificial networks that were generated by the standard stochastic blockmodel (SBM) \citep{holland1983stochastic}. In fact, the SBM can produce flexible and challenging synthetic networks with a wide variety of network structures. And the strength of network structures is easily controlled. After generating artificial networks, the attributes related to communities were produced by 0-1 distribution $ Bin(1, p) $, where $ p $ measures how well a community matches attributes. The larger the values of $ p $, the stronger the dependency between attributes and communities. Here, assume that each community has a strong dependency with 10 attributes ($ p > 0.1 $), but has little relationship with the remaining attributes ($ p=0.1 $).
	
\begin{figure}[htbp]
	\centering
	\includegraphics[width=0.9\textwidth]{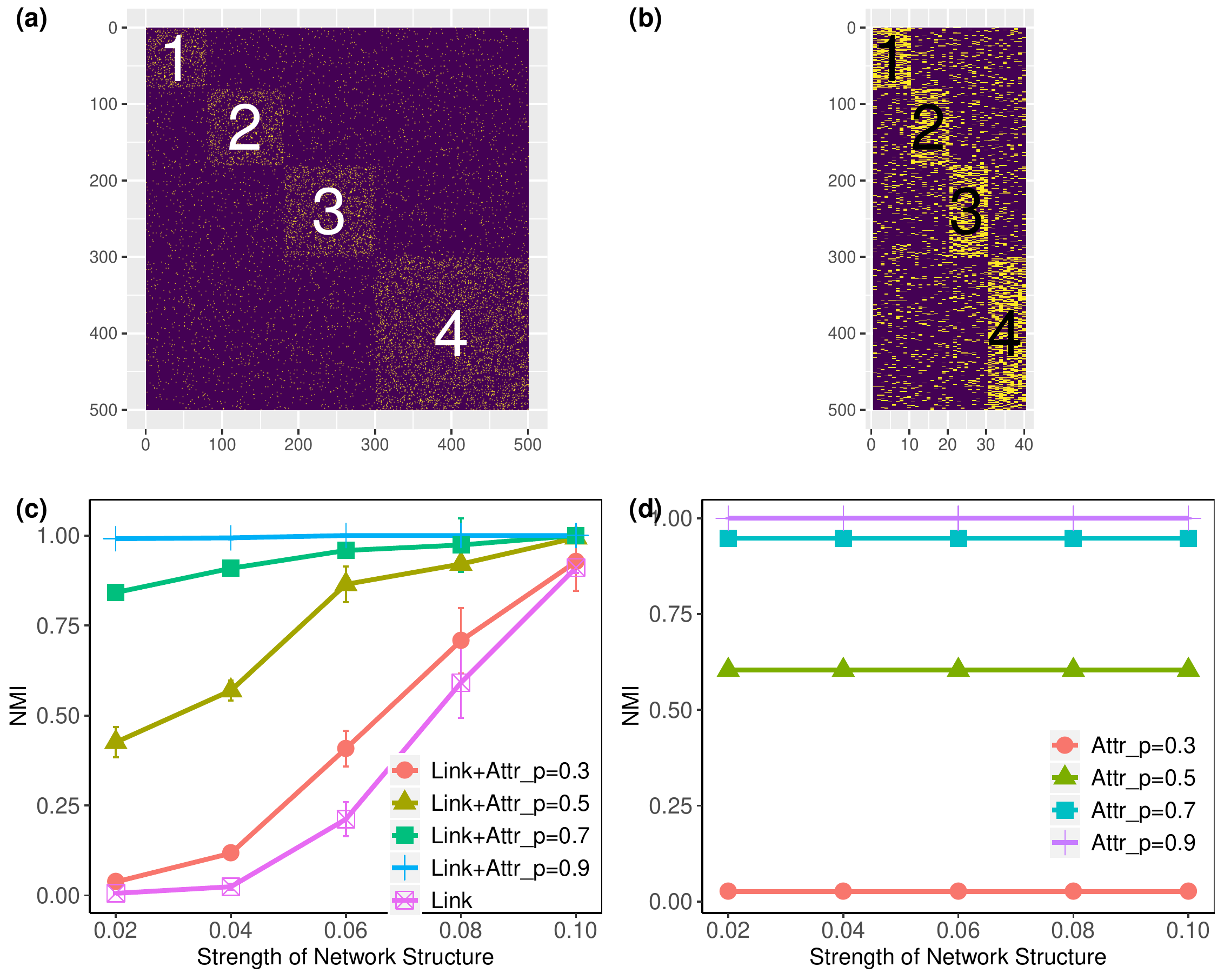}
	\caption{The performance of GNAN on networks with traditional communities. In this group of studies, $ \lambda \equiv 0.02 $, $ \omega = 0.02, 0.04, 0.06, 0.08, 0.10$, the corresponding network structures are gradually clear, the sizes of 4 communities are 80, 100, 120, and 200, respectively. (a) A tested network with community structure ($ \lambda  = 0.02,\;\omega  = 0.10 $)  \textemdash a set of communities with dense internal connections and sparse external ones, ``1, 2, 3", and ``4" are 4 communities, respectively. (b) An attribute matrix used in this group of tests ($ p=0.7 $), 10 attributes are matched to each community. (c) Both links and attributes were used (Link+Attr), only links were used (Link). (d) Only attributes were used (Attr). ``p=0.3, 0.5, 0.7, 0.9" measure the strength that a community matches attributes.}\label{fig2}
   \end{figure}

	\textbf{Community structure.} Suppose the parameter generating network structures in SBM is $ \omega $ which has the following particular form
	\[
	\omega ^{{\rm{planted}}}  = \left( \begin{array}{l}
	\omega \;\;\lambda \;\;\lambda \;\;\lambda  \\ 
	\lambda \;\;\omega \;\;\lambda \;\;\lambda  \\ 
	\lambda \;\;\lambda \;\;\omega \;\;\lambda  \\ 
	\lambda \;\;\lambda \;\;\lambda \;\;\omega  \\ 
	\end{array} \right),  \eqno(9)
	\]
	where $ \omega \geq \lambda $. The smaller the difference between $ \omega $ and $ \lambda $, the vaguer the network structures. $ \omega = \lambda $ means a fully random network with no group structure. The 10 attributes that have a strong relationship with each community were produced by $ Bin(1, p) $ with $ p=0.3, 0.5, 0.7 $, or 0.9. The results were shown in FIG.\ref{fig2}.

	From FIG.\ref{fig2} (c), we can easily see that (1) community detection results with both topology and attributes are better than ones without attributes; (2) the clearer the network structure, the better the detection effect; (3) the closer the relationship between the community and the attribute, the better the detection accuracy. These conclusions are consistent with our intuition, which shows the node-attribute information can help to improve community detection. Especially when the network has no community structure ($ \omega = \lambda =0.02 $), the results of community detection are entirely determined by attributes (i.e., $ \phi_{rk} $). Similar conclusions are shown in FIG.\ref{fig2} (d), where links are not considered, community detection is controlled by attributes ($ \phi_{rk} $), and the larger the $ p $, the better the community detection.
	
\begin{figure}[htbp]
		\centering
		\includegraphics[width=0.9\textwidth]{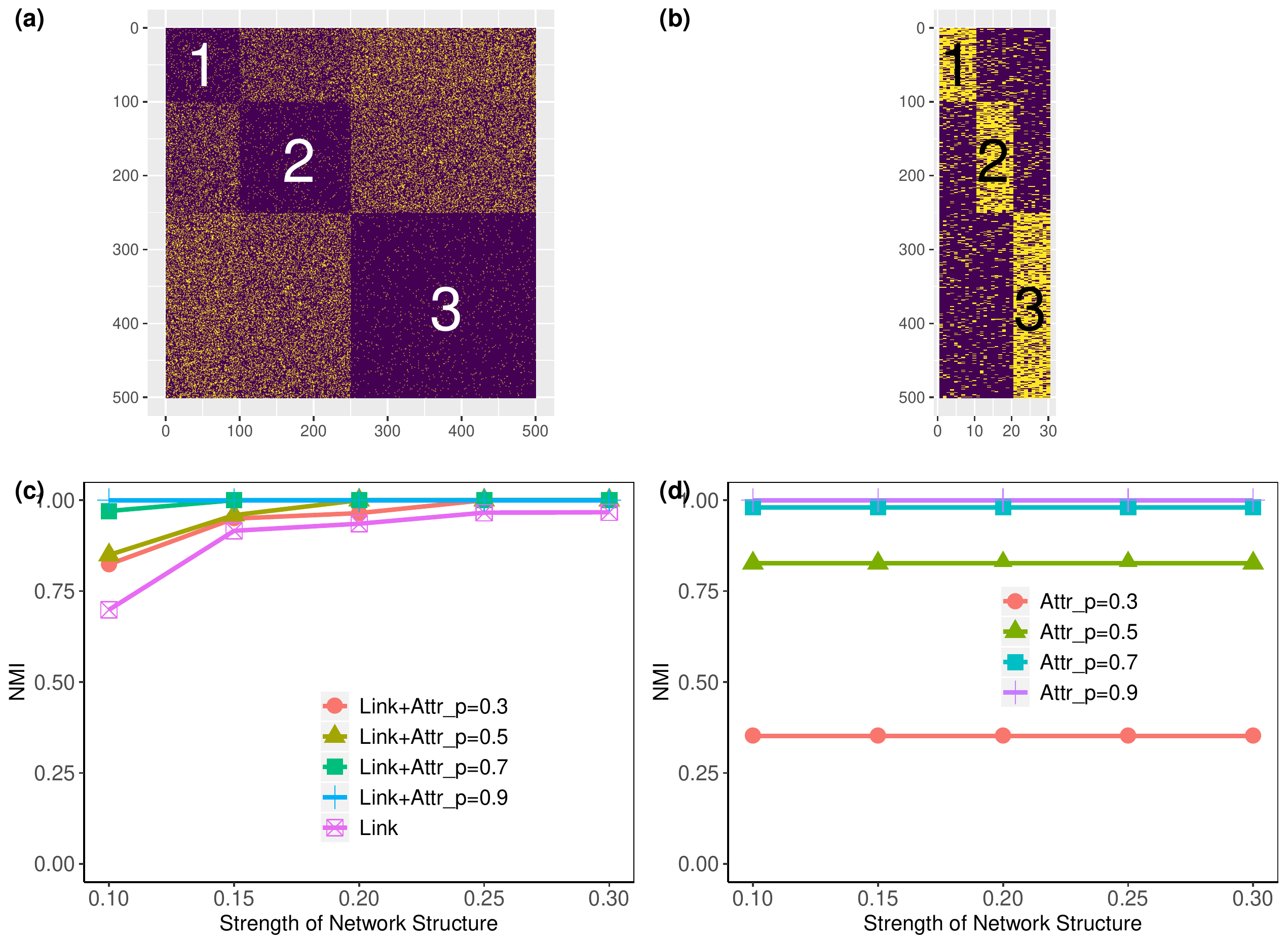}
		\caption{The performance of GNAN on networks with disassortative communities. In this group of studies, $ \lambda_1 = 0.10, 0.15, 0.20, 0.25, 0.30$, the corresponding network structures are gradually clear, the sizes of 3 communities are 100, 150, and 250, respectively. (a) An illustration with disassortative structure ($ \lambda_1  = 0.20 $), in which nodes have most of their connections outside their community. (b) An attribute matrix ($ p=0.7 $), 10 attributes are matched to each community. (c) Both links and attributes were used (Link+Attr), only links were used (Link). (d) Only attributes were used (Attr). Other symbols are the same as the ones in FIG. \ref{fig2}.}\label{fig3}
	\end{figure}

	\textbf{Disassortative structure.} A group of five networks including disassortative structure was generated by SBM with parameter
   \[
   \omega ^{{\rm{planted}}}  = \left( {\begin{array}{*{20}c}
	{0.05} & {\lambda _1 } & {\lambda _1  + 0.1}  \\
	{\lambda _1 } & {0.03} & {\lambda _1  + 0.05}  \\
	{\lambda _1  + 0.1} & {\lambda _1  + 0.05} & {0.02}  \\
	\end{array}} \right) \eqno(10)
   \]
		where $ \lambda_1 > 0.05 $. The smaller the $ \lambda_1 $, the vaguer the network structures. The results were shown in FIG. \ref{fig3}.
		
	As in FIG. \ref{fig3} (c-d), the same conclusion as in FIG. \ref{fig2} can be derived. The combination of attribute information and topological information improves the accuracy of community discovery. These conclusions are in line with expectations. 
	
\begin{figure}[htb]
		\centering
		\includegraphics[width=0.9\textwidth]{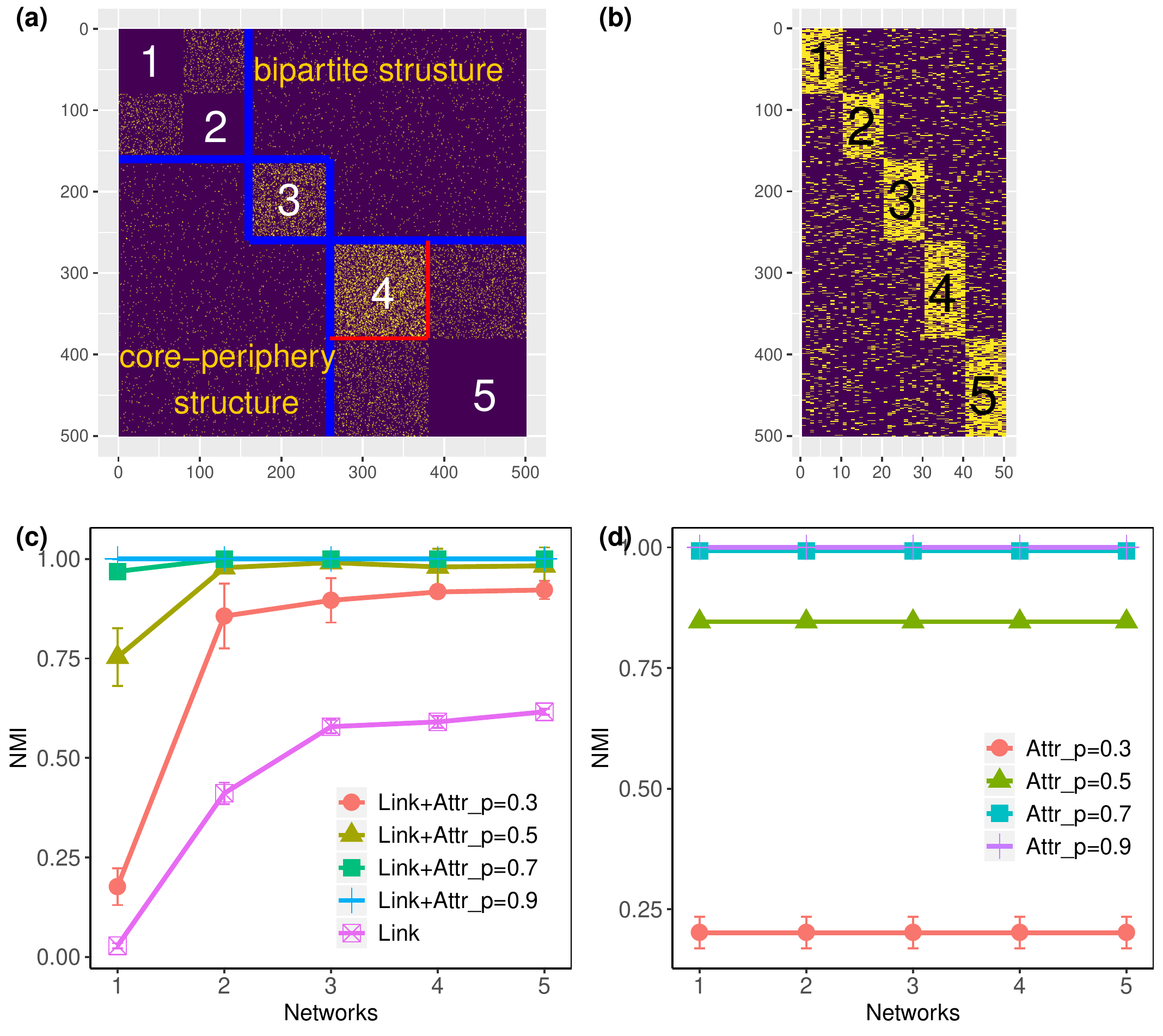}
		\caption{The performance of GNAN on networks with mixture structures (bipartite structure, community structure, and core-periphery structure). The parameters corresponding to the 5 networks are as follows: $  \lambda \equiv 0.02 $, and network 1: $ \omega_1 = \omega_2 =  \omega_3 =  \omega_4 = 0.05$; network 2: $ \omega_1 = \omega_2 =  \omega_3 =  \omega_4 = 0.1$; network 3: $ \omega_1 = 0.1, \omega_2 = 0.2, \omega_3 = 0.4, \omega_4 = 0.1$; network 4: $ \omega_1 = 0.1, \omega_2 = 0.3, \omega_3 = 0.4, \omega_4 = 0.1$; network 5: $ \omega_1 = 0.1, \omega_2 = 0.5, \omega_3 = 0.4, \omega_4 = 0.1$. (a) An illustration with a mixture structure (network 3). Other symbols are the same as the ones in FIG.\ref{fig2}.} \label{fig4}
	\end{figure}

	\textbf{Mixture structure.} A group of five networks including bipartite structure, community structure and core-periphery structure was generated by SBM with parameter
	\[
	\omega ^{{\rm{planted}}}  = \left( \begin{array}{l}
	0\;\;\omega _1 \;\;\lambda \;\;\lambda \;\;\lambda  \\ 
	\omega _1 \;\;0\;\;\;\lambda \;\;\lambda \;\;\lambda  \\ 
	\lambda \;\;\lambda \;\;\omega _2 \;\;\lambda \;\;\lambda  \\ 
	\lambda \;\;\lambda \;\;\lambda \;\;\omega _3 \;\;\omega _4  \\ 
	\lambda \;\;\lambda \;\;\lambda \;\;\omega _4 \;\;0 \\ 
	\end{array} \right)  \eqno(11)
	\]
	to evaluate the performance of our new model. The 10 attributes that have a strong relationship with each community were produced by $ Bin(1, p) $ with $ p=0.3, 0.5, 0.7 $, or 0.9. The results were shown in FIG. \ref{fig4}.

    From FIG. \ref{fig4} (c-d), on the networks with mixture structure, the performance of GNAN is almost the same as the performance on the network with community structure (FIG. \ref{fig2}) and disassortative structure (FIG. \ref{fig3}). These phenomena in both FIG. \ref{fig2} and FIG. \ref{fig3} mean that our new model GNAN can effectively use the node-attribute information to improve the community detection on networks with generalized structures.
    
    The above experiments were mainly designed to test community detection. In order to evaluate whether our model GNAN can automatically discover important node attributes and ignore attributes without important information, we designed the following group of experiments.
    
    \textbf{The ability to discover attribute information.} The network used here was the one in FIG. \ref{fig2} ($ \lambda = 0.02, \omega = 0.10 $). The attributes corresponding to each community were designed as follows.
    
    \begin{table}[H]
    	\centering
    	\caption{The dependency between communities and attributes. ``-" represents that the 10 attributes were generated by $ Bin(1, p) $ with $ p=0.1 $, which means that these 10 attributes are noisy for the corresponding community. Note that attributes 31-40 are noisy for all communities.}
    	\resizebox{0.7\textwidth}{!}{
    		\begin{tabular}{c c c c c}
    			\hline
    			Attributes  & 1-10 &  11-20 & 21-30  & 31-40  \\
    			\hline
    			Community\_1& Strong ($ p=0.9 $) & Strong ($ p=0.9 $) & - &- \\
    			Community\_2 & Strong ($ p=0.9 $)  & Strong ($ p=0.9 $) & - & -  \\
    			Community\_3 & - & - & Strong ($ p=0.7 $) & - \\
    			Community\_4 & - & - & Strong ($ p=0.7 $) & -  \\
    			\hline
    		\end{tabular}%
    	}
    	\label{tab1}%
    \end{table}%

\begin{figure}[htbp]
    	\centering
    	\includegraphics[width=0.9\textwidth]{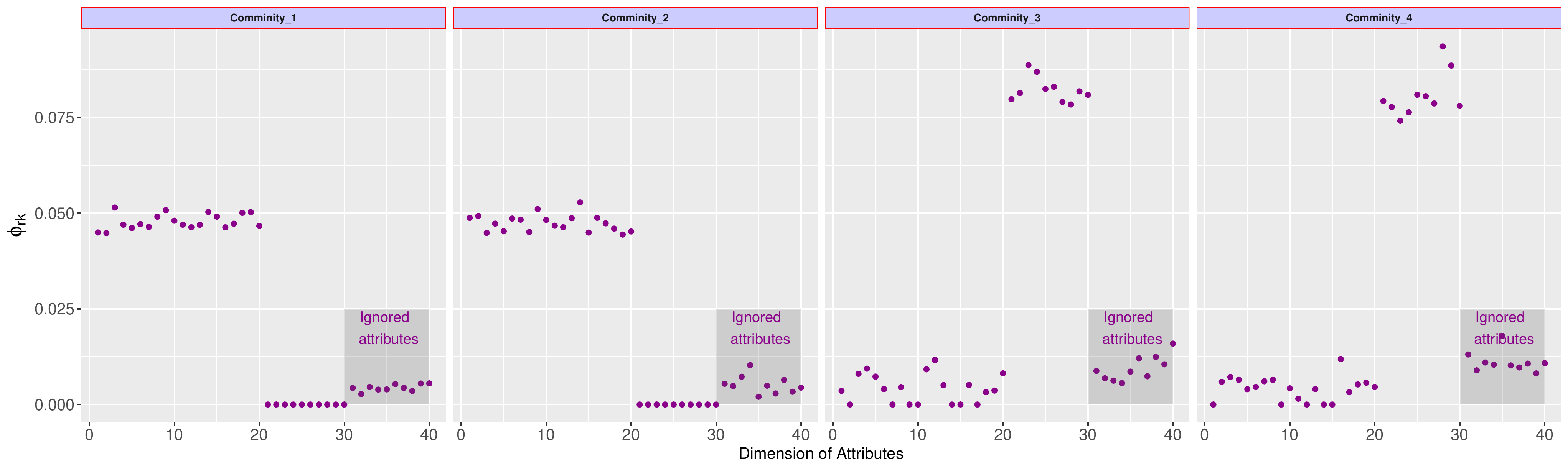}
    	\caption{The inferred $ \phi_{rk} $ which characters the ability to find attribute information. ``Ignored attributes" means that the 10 attributes are noisy for all communities such that they can be ignored when considering the importance of attributes.}\label{fig5}
    \end{figure}

    Using this node-attribute network, the inferred dependencies by our model GNAN were shown in FIG. \ref{fig5}.
    
    As can be seen in FIG. \ref{fig5}, for community\_1, the values of $ \phi_{rk} $ from 1 to 20 are significantly larger than the rest, which means the corresponding 20 attributes are discovered by the GNAN. This conclusion is in line with the truth in TAB. \ref{tab1}. For the other three communities, the same conclusions are derived. Note that attributes 31-40 are noisy for all communities in TAB. \ref{tab1}, the corresponding values of $ \phi_{rk} $ are always small compared to the ones of strongly dependent attributes. These phenomena show that the model GNAN has the ability to find important attributes, which is beneficial to the semantic interpretation of communities in practice. We will see this conclusion in the following case study.

	\subsection{Semantic interpretation of communities: an example about the friendship between Lazega’s lawyers}
	In this section, through a small real-world network, we revealed what the main characteristics of each community are and tried to explain why they became friends. This data set includes 71 attorneys and 575 links and comes from a network study of corporate law partnership that was carried out in a Northeastern US corporate law firm, referred to as SG \& R, 1988-1991 in New England. The dataset has various members' attributes as follows:
   
   \begin{table}[htbp]
   	\centering
   	%\caption{}
   	\resizebox{0.8\textwidth}{!}{
   		\begin{tabular}{l  l}
   			\textbullet $ \; $seniority (range: 1-71) & $ \; $  \textbullet $ \; $status (1=partner; 2=associate) \\
   			\textbullet $ \; $gender (1=man; 2=woman) & $ \; $ \textbullet $ \; $office (1=Boston; 2=Hartford; 3=Providence)  \\
   			\textbullet $ \; $years with the firm (range: 1-32) & $ \; $ \textbullet $ \; $age (range: 26-67) \\
   			\textbullet $ \; $practice (1=litigation; 2=corporate) & $ \; $ \textbullet $ \; $law school (1: harvard, yale; 2: ucon; 3: other)  \\
   		\end{tabular}%
   	}
   	\label{tab0}%
   \end{table}%
   where office means the office in which they work and other names of items are self-explanatory. Because the number of communities is not given in advance in this friendship network, by maximizing the modularity measure (Q) over all possible partitions (Q$ _{max} $=0.4088), 4 ground-truth communities were obtained, where 2 isolated points (NO.44 and NO.47) were removed. The communities were shown in FIG. \ref{fig5}, where a color represents a community. Here, the attribute variables ``age" and ``years with the firm" were discretized as shown in TAB. \ref{tab2}.
   
  \begin{table}[H]
  	\centering
  	\caption{The inferred $ \phi_{rk} $ for each community, and the ones that greater than 0.1 were in bold.}
  	\begin{tabular}{c|ccccc}
  		\hline
  		\multicolumn{2}{c}{Attributes} & Comminity\_1 & Comminity\_2 & Comminity\_3 & Comminity\_4 \\
  		\hline
  		\multirow{2}[2]{*}{status } & partner & \textbf{0.1479 } & 0.0990  & 0.0000  & 0.0443  \\
  		& associate & 0.0000  & 0.0338  & \textbf{0.1609 } & 0.0614  \\
  		\hline
  		\multirow{2}[2]{*}{gender} & man   & \textbf{0.1361 } & \textbf{0.1334 } & 0.0817  & 0.0682  \\
  		& woman & 0.0000  & 0.0171  & 0.0560  & 0.0906  \\
  		\hline
  		\multirow{3}[2]{*}{office} & Boston & \textbf{0.1232 } & 0.0000  & \textbf{0.1321 } & \textbf{0.1065 } \\
  		& Hartford & 0.0000  & \textbf{0.2002 } & 0.0000  & 0.0000  \\
  		& Providence & 0.0063  & 0.0000  & 0.0000  & 0.0138  \\
  		\hline
  		\multirow{3}[2]{*}{age} & $ \leq $35  & 0.0000  & 0.0684  & \textbf{0.1111 } & 0.0000  \\
  		& 36-45 & 0.0000  & 0.0411  & 0.0366  & \textbf{0.1704 } \\
  		& $ \geq $ 46  & \textbf{0.1453 } & 0.0000  & 0.0000  & 0.0000  \\
  		\hline
  		\multirow{3}[2]{*}{years with the firm} & 1-4   & 0.0000  & 0.0499  & \textbf{0.1414 } & 0.0000  \\
  		& 5-9   & 0.0000  & 0.0301  & 0.0000  & \textbf{0.1540 } \\
  		& $ \geq $10  & \textbf{0.1554 } & 0.0359  & 0.0000  & 0.0000  \\
  		\hline
  		\multirow{2}[2]{*}{practice} & litigation & 0.0724  & 0.0771  & 0.0923  & 0.0924  \\
  		& corporate & 0.0659  & 0.0798  & 0.0434  & 0.0576  \\
  		\hline
  		\multirow{3}[2]{*}{law school} & harvard or yale & 0.0781  & 0.0000  & 0.0000  & 0.0227  \\
  		& ucon  & 0.0291  & 0.0757  & 0.0836  & 0.0421  \\
  		& other & 0.0404  & 0.0584  & 0.0608  & 0.0759  \\
  		\hline
  	\end{tabular}%
  	\label{tab2}%
  \end{table}%

	\begin{figure}[htb]
		\centering
		\includegraphics[width=0.8\textwidth]{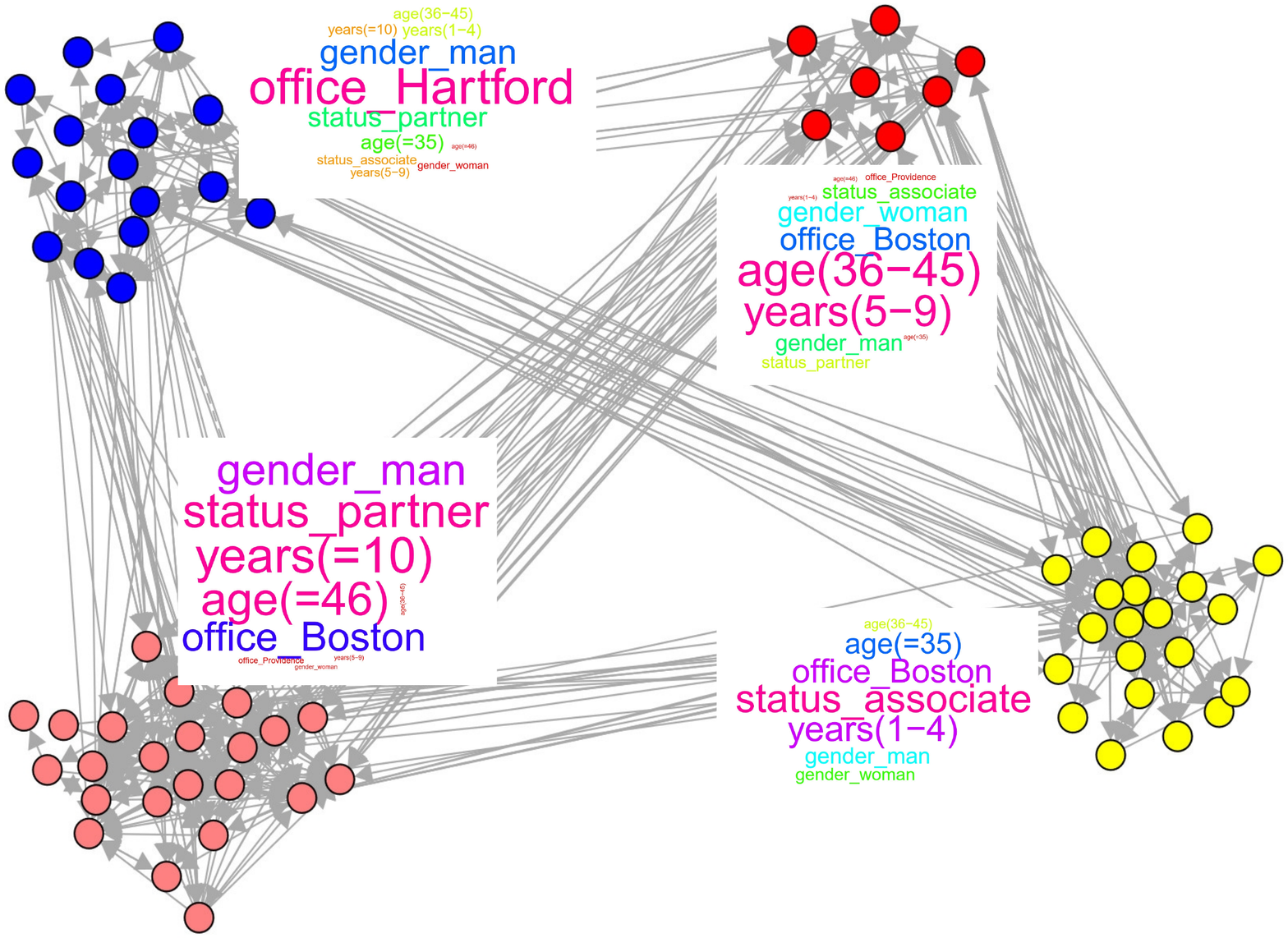}
		\caption{A real-world network with 71 attorneys and 575 links between them. Each node has 18 attributes that belong to 6 classes as shown in TAB. \ref{tab2}. By maximizing the modularity measure (Q) over all possible partitions (Q$ _{max} $=0.4088), 4 ground-truth communities were obtained, where 2 isolated points (NO.44 and NO.47) were removed. Different colors represent different communities. The semantic interpretation inferred by the new method was shown near the corresponding community.}\label{fig6}
	\end{figure}
	
	As can be seen in TAB. \ref{tab2}, taking 0.1 as a boundary, Community\_1 has the most attributes (5 attributes), while Community\_2 has the least (2 attributes), which shows that there may be many reasons why people become friends, or they may not need many reasons. From the perspective of attributes, the attribute variables ``practice " and ``law school" are not important to every community, on the contrary, ``office" is important to each community, which means that distance (or geographic location) is one of the important factors that affect people's friendships, while the specific works they are engaged in and the schools they used to attend are not so important for friendships in this network. These conclusions show the ability of our model GNAN to learn attributes automatically, which helps the semantic interpretation of each community. A visual semantic explanation was shown in FIG. \ref{fig5}.
	
    From TAB. \ref{tab2} and FIG. \ref{fig5}, we can try to explain why people build their friendships. From the workplace, each community seems to be formed like this: firstly, people in office Hartford formed a community (Community\_2). Then, the rest people (mainly office Boston) were divided into three communities: Community\_4 contains middle-aged persons (36-45 years old) who have worked in the law firm for between 5 and 10 years; Community\_3 is mainly composed of young people (less than 35 years old) who have worked for no more than 5 years; people in Community\_1 seem to be friends because they have been in the company for a long time ($ \geq 10 $), have almost the same status (partner), are old ($ \geq 46 $), and work in the same office (Boston).

	\subsection{Comparison of our model GNAN with other state-of-the-art models} 
	We showed our model GNAN for both community detection and the ability to find important attributes in the above experiments. Next, we would compare the new model GNAN with other 4 state-of-the-art models: PSB\_PG, NEMBP, BNPA, and PPSB\_DC, which are all probabilistic generative models and can detect generalized structure. The results were shown in the following TAB. \ref{tab3} and TAB. \ref{tab4}.
	
	\begin{table}[htbp]
		\centering
		\caption{Comparison results (mean $ \pm $ error) on artificial networks with a range of network structures. The best results were in bold. The networks $ \omega $0.06 and $ \omega $0.04 are the ones $ \omega $=0.06 and $ \omega $=0.04 in FIG. \ref{fig2}. ``m3" and ``m4" mean the third and fourth networks in FIG. \ref{fig4}. ``d0.1" means the network $ \lambda_1=0.1 $ in FIG. \ref{fig3}. ``cp" means a network with core-periphery structure in FIG. \ref{fig4}, and parameter $ \omega_3=0.12,  \omega_4=0.1$. ``$ \omega $0.1\_noisyAttr" is the network with noisy attributes in TAB. \ref{tab1}.}
		\resizebox{\textwidth}{!}{
			\begin{tabular}{llccccc}
				\hline
				Network & Structure & GANA  & PSB\_PG & NEMBP & BNPA  & PPSB\_DC \\
				\hline
				SBM\_$ \omega $0.06\_Attr0.9 & Community & \textbf{1.0000$ \pm $0.0000} & \textbf{1.0000$ \pm $0.0000} & 0.9960$ \pm $0.0000 & \textbf{1.0000$ \pm $0.0000} & 0.9485$ \pm $0.0728 \\
				SBM\_$ \omega $0.04\_Attr0.9 & Community & \textbf{0.9922$ \pm $0.0048} & 0.9844$ \pm $0.0000 & 0.9127$ \pm $0.0947 & 0.9901$ \pm $0.0206 & 0.9382$ \pm $0.0564 \\
				SBM\_m3\_Attr0.5 & Mixture & \textbf{0.9908$ \pm $0.0035} & 0.9378$ \pm $0.0782 & 0.9527$ \pm $0.0883 & 0.7569$ \pm $0.0175 & 0.5526$ \pm $0.0771 \\
				SBM\_m4\_Attr0.5 & Mixture & 0.9798$ \pm $0.0459 & \textbf{0.9963$ \pm $0.0102} & 0.9777$ \pm $0.0629 & 0.86854$ \pm $0.0000 & 0.7577$ \pm $0.0077 \\
				SBM\_d0.1\_Attr0.5 & Disassortative & \textbf{0.9416$ \pm $0.0909} & 0.9285$ \pm $0.1085 & 0.8046$ \pm $0.0000 & 0.8020$ \pm $0.0000 & 0.6428$ \pm $0.0160 \\
				SBM\_d0.1\_Attr0.3 & Disassortative & \textbf{0.8236$ \pm $0.0855} & 0.7746$ \pm $0.1567 & 0.8052$ \pm $0.0000 & 0.8026$ \pm $0.0000 & 0.3791$ \pm $0.0000 \\
				SBM\_cp\_Attr0.5 & Core-periphery & 0.9393$ \pm $0.0000 & 0.9519$ \pm $0.0000 & \textbf{1.0000$ \pm $0.0000} & 0.9426$ \pm $0.0027 & 0.6271$ \pm $0.0106 \\
				SBM\_cp\_Attr0.4 & Core-periphery & \textbf{0.9162$ \pm $0.0000} & 0.9024$ \pm $0.0049 & 0.8006$ \pm $0.0210 & 0.9144$ \pm $0.0017 & 0.3726$ \pm $0.0062 \\
				SBM\_$ \omega $0.1\_noisyAttr & Community & \textbf{0.9443$ \pm $0.0750} & 0.8295$ \pm $0.0684 & 0.9348$ \pm $0.0816 & 0.8970$ \pm $0.0000 & 0.6608$ \pm $0.0000 \\
				\hline
			\end{tabular}%
		}
		\label{tab3}%
	\end{table}%
	
	As in TAB. \ref{tab3}, on the networks with traditional structure, the performance of these five methods is good, especially the methods GANA, PSB\_PG, and BNPA. On the networks with mixture structure, the methods GANA, PSB\_PG, and BNPA are superior to the rest 2 methods. On the networks with disassortative structure and core-periphery structure, all these methods except PPSB\_DC are highly efficacious. Simply speaking, on most of the tested networks, the community detection quality of the new method is the best in all methods. On the contrary, the performance of the method PPSB\_DC is the worst.

	\begin{table}[htbp]
		\centering
		\caption{Comparison results (mean $ \pm $ error) on real-world networks. The best results were in bold.}
		\resizebox{\textwidth}{!}{
		\begin{tabular}{llccccccccc}
			\hline
			Network & N & M & K & C & Structure & GANA  & PSB\_PG & NEMBP & BNPA  & PPSB\_DC \\
			\hline
			Cornell & 195 & 304 & 1703 & 5 & Community & \textbf{0.3505$ \pm $0.0533} & 0.3115$ \pm $0.0576 & 0.1890$ \pm $0.0416 & 0.2211$ \pm $0.0077 & 0.1257$ \pm $0.0110 \\
			Texas & 187 & 328 & 1703 & 5 & Community & \textbf{0.3226$ \pm $0.0280} & 0.3072$ \pm $0.0362 & 0.3093$ \pm $0.0222 & 0.1922$ \pm $0.0296 & 0.1576$ \pm $0.0101 \\
			Washington & 230 & 446 & 1703 & 5 & Community & \textbf{0.3433$ \pm $0.0414} & 0.3013$ \pm $0.0323 & 0.2085$ \pm $0.0407 & 0.1697$ \pm $0.0158 & 0.2818$ \pm $0.0444 \\
			Wisconsin & 265 & 530 & 1703 & 5 & Community & \textbf{0.4200$ \pm $0.0258} & 0.3729$ \pm $0.0279 & 0.3004$ \pm $0.0427 & 0.2696$ \pm $0.0126 & 0.2272$ \pm $0.0407 \\
			Cora  & 2708 & 5429 & 1433 & 7 & Mixture & 0.3594$ \pm $0.0373 & 0.3442$ \pm $0.0382 & 0.4188$ \pm $0.0255 & \textbf{0.4780$ \pm $0.0303} & 0.4659$ \pm $0.0090 \\
			Citeseer & 3312 & 4723 & 3703 & 6 & Mixture & 0.2606$ \pm $0.0296 & 0.2543$ \pm $0.0364 & 0.2325$ \pm $0.0192 & 0.2958$ \pm $0.0321 & \textbf{0.3753$ \pm $0.0382} \\
			\hline
		\end{tabular}%
	}
		\label{tab4}%
	\end{table}%
	
From TAB. \ref{tab4}, on the first four networks with community structure, the methods GANA and PSB\_PG are superior to the other 2 methods. Instead, on networks with mixture structure (Cora and Citeseer), the methods BNPA and PPSB\_DC are superior to the rest four methods. The performance of various methods on the real and artificial networks is not completely consistent, which means that there is a gap between the computer-generated network structure and the real network structure. However, whether on synthetic networks or real networks, the community detection quality of the new method GANA is competitive with the other state-of-the-art methods. 

\begin{figure}[htbp]
	\centering
	\includegraphics[width=0.8\textwidth]{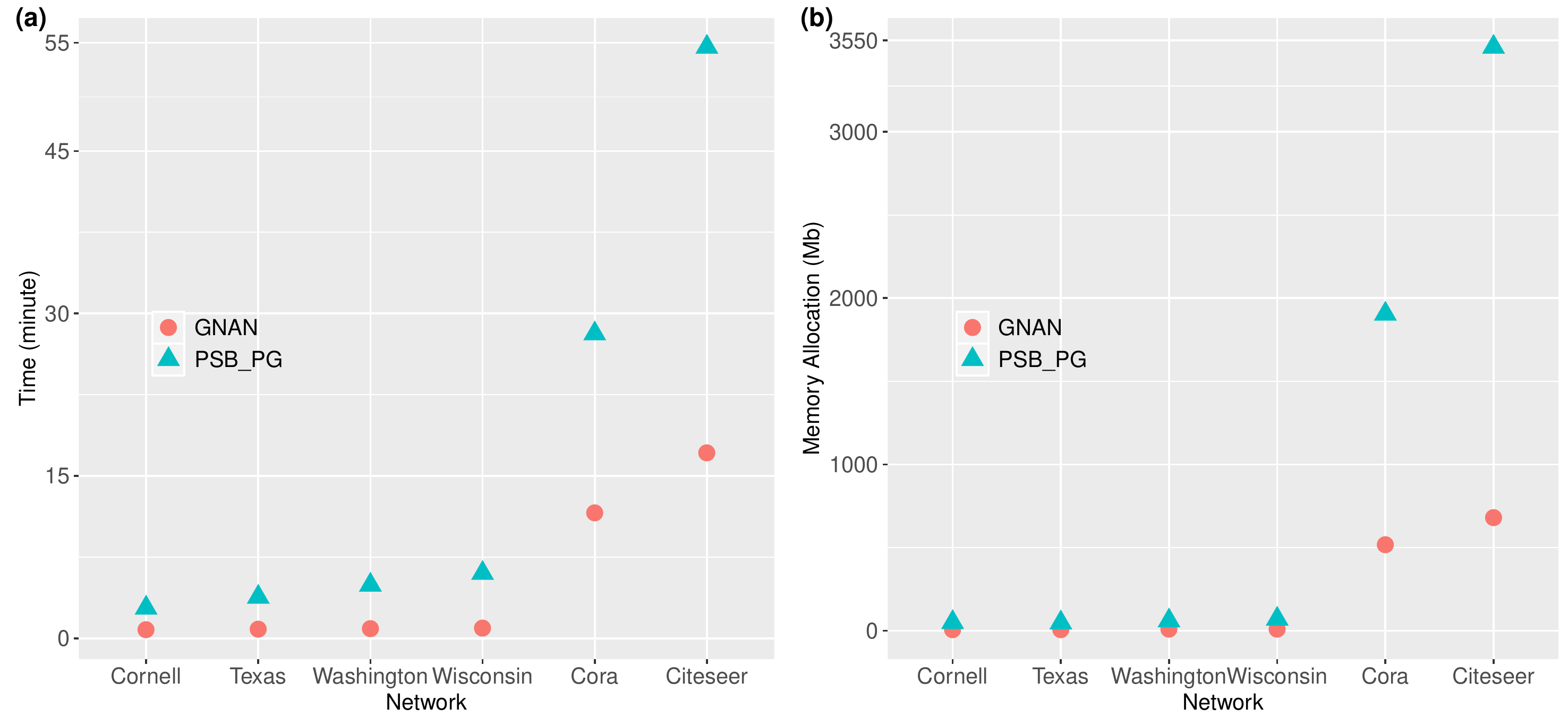}
	\caption{Comparison of the complexity. In the two algorithms compared, the running parameters are the same as the original. The time and memory here were obtained when the corresponding algorithm reaches its stop condition. }\label{fig7}
\end{figure}

In addition, the complexity of the PSB\_PG mentioned above is higher than that of our new method. We showed the comparison results on real networks (see FIG. \ref{fig7}). From FIG. \ref{fig7}, the new algorithm GNAN is superior to the algorithm PSB\_PG, especially on the latter two networks.

\section{Conclusions}
In summary, based on the connectivity behavior of nodes and homogeneity of attributes, we have developed a generative node-attribute network model GNAN that combines topological information and attribute information. The major contributions: (1) The new model GNAN can detect a range of network structures, by experiments in Sect.IV, we have shown this feature. (2) The node attributes that match the true community assignments of nodes can be automatically learned by our model. We have designed an experiment to test this ability of the model GNAN. By using this ability, a case study has been provided to show the semantic interpretability of communities. (3) The new model detects communities more accurately than the model that only uses topology information. Experiments on both synthetic and real-world networks have shown that the new model is competitive with other state-of-the-art models.

\section*{Acknowledgements}
This work is supported in part by the National Natural Science Foundation of China (granted No. 61876016), National Key R \& D Program of China (No. 2018AAA0100302), the Higher Education Innovation Ability Improvement Project in Gansu Province (No. 2020-098A), and the Research Project at Tianshui Normal University (No. CXJ2020-28). The authors thank the anonymous reviewers for their constructive comments.

   %\appendix*
   \section*{Appendix A}
	Note that the constraint on $ \tau_{ir} $: $ \sum\limits_{r = 1}^C {\tau _{ir} }  = 1 $, and ignoring the constants not related to T, we have
	\[
	\tilde L\left( {\rm T} \right) = \sum\limits_{ijr} {a_{ij} q_{ij,r} \log \tau _{ir} }  - \sum\limits_{ijr} {\tau _{ir} \theta _{rj} }  + \sum\limits_{ikr} {x_{ik} h_{ik,r} \log \tau _{ir} }  - \sum\limits_{ikr} {\tau _{ir} \varphi _{rk} }  + \sum\limits_i {\lambda _i \left( {\sum\limits_r {\tau _{ir} }  - 1} \right)}. \eqno(A1)
	\]
	Taking the first-order partial derivative of the Lagrangian $\tilde L\left( {\rm T} \right)$ with respect to $\tau_{ir}$, we have
	\[
	\frac{{\partial \tilde L\left( {\rm T} \right)}}{{\partial \tau _{ir} }} = \frac{{\sum\limits_j {a_{ij} q_{ij,r} } }}{{\tau _{ir} }} - \sum\limits_j {\theta _{rj} }  + \frac{{\sum\limits_k {x_{ik} h_{ik,r} } }}{{\tau _{ir} }} - \sum\limits_k {\phi _{rk} }  + \lambda _i.  \eqno(A2)
	\]
	
	Set $ \frac{{\partial \tilde L\left( {\rm T} \right)}}{{\partial \tau _{ir} }} = 0 $, we have
	\[
	\frac{{\sum\limits_j {a_{ij} q_{ij,r} }  + \sum\limits_k {x_{ik} h_{ik,r} } }}{{\tau _{ir} }} - 2 + \lambda _i {\rm{ = 0}}, \eqno(A3)
	\]
	
	\[
	\sum\limits_j {a_{ij} q_{ij,r} }  + \sum\limits_k {x_{ik} h_{ik,r} }  - 2\tau _{ir}  + \lambda _i \tau _{ir} {\rm{ = 0}},  \eqno(A4)
	\]
	
	\[
	\sum\limits_{jr} {a_{ij} q_{ij,r} }  + \sum\limits_{kr} {x_{ik} h_{ik,r} }  - 2\sum\limits_r {\tau _{ir} }  + \lambda _i \sum\limits_r {\tau _{ir} } {\rm{ = 0}}. \eqno(A5)
	\]
	
	By $ (A3) $ and $ (A5) $, we have
	\[
	\tau _{ir}  = \frac{{\sum\limits_j {a_{ij} q_{ij,r} }  + \sum\limits_k {x_{ik} h_{ik,r} } }}{{\sum\limits_{jr} {a_{ij} q_{ij,r} }  + \sum\limits_{kr} {x_{ik} h_{ik,r} } }}. \eqno(A6) 
	\]
	
	Similarly, by Eq.(5) and the constraint $ \sum\limits_{j = 1}^N {\theta _{rj} }  = 1 $, we have 
	\[
	\tilde L\left( \Theta  \right) = \sum\limits_{ijr} {a_{ij} q_{ij,r} \log \theta _{rj}  - } \sum\limits_{ijr} {\tau _{ir} \theta _{rj} }  + \sum\limits_r {\eta _r \left( {\sum\limits_j {\theta _{rj} }  - 1} \right)}. \eqno(A7)
	\]
	
	Taking the first order partial derivative of the Lagrangian $\tilde L\left( \Theta \right)$ with respect to $\theta _{rj}$ and set it to be zero, we have
	\[
	\frac{{\sum\limits_i {a_{ij} q_{ij,r} } }}{{\theta _{rj} }} - \sum\limits_i {\tau _{ir} }  + \eta _r  = 0.   \eqno(A8)
	\]
	
	\[
	\sum\limits_{ij} {a_{ij} q_{ij,r} }  - \sum\limits_j {\theta _{rj} } \sum\limits_i {\tau _{ir} }  + \sum\limits_j {\eta _r \theta _{rj} }  = 0.  \eqno(A9)
	\]
	
	By $ (A8) $ and $ (A9) $, we have
	\[
	\theta _{rj}  = \frac{{\sum\limits_i {a_{ij} q_{ij,r} } }}{{\sum\limits_{ij} {a_{ij} q_{ij,r} } }}.   \eqno(A10)
	\]
	
	Finally, using the condition $ \sum\limits_{k = 1}^K {\phi _{rk} }  = 1 $ and  Eq.(5), we have
	\[
	\tilde L\left( \Phi  \right) = \sum\limits_{ikr} {x_{ik} h_{ik,r} \log \phi _{rk} }  - \sum\limits_{ikr} {\tau _{ir} \phi _{rk} }  + \sum\limits_r {\rho _r \left( {\sum\limits_k {\phi _{rk} }  - 1} \right)}.   \eqno(A11)
	\]
	
	Taking the first derivative of the Lagrangian $\tilde L\left( \Phi \right)$ with respect to $\phi _{rk}$ and set it to be zero, we have
	\[
	\frac{{\sum\limits_i {x_{ik} h_{ik,r} } }}{{\phi _{rk} }} - \sum\limits_i {\tau _{ir} }  + \rho _r  = 0.  \eqno(A12)
	\]
	
	\[
	\sum\limits_{ik} {x_{ik} h_{ik,r} }  - \sum\limits_{ik} {\tau _{ir} \phi _{rk} }  + \sum\limits_k {\eta _r \phi _{rk} }  = 0.  \eqno(A13)
	\]
	
	By $ (A12) $ and $ (A13) $, we have
	\[
	\phi _{rk}  = \frac{{\sum\limits_i {x_{ik} h_{ik,r} } }}{{\sum\limits_{ik} {x_{ik} h_{ik,r} } }}.   \eqno(A14)
	\]

%\section*{References}

\bibliography{nmm20200322}% Produces the bibliography via BibTeX.
	
%\begin{thebibliography}{00}	

%	\bibitem{C} M. E. J. Newman, Networks, Oxford University Press, Oxford, 2nd Edition, 2018.
	
%\end{thebibliography}

\end{document}